\documentclass[aps,prl,twocolumn,showpacs,preprintnumbers]{revtex4}
\usepackage{graphicx}

\begin{document}

\title{On the work distribution for the adiabatic compression of a dilute classical gas}

\author{Gavin E. Crooks }
\email{gecrooks@lbl.gov}
\affiliation{Physical Biosciences Division, Lawrence Berkeley National Laboratory, Berkeley, California 94720}
\author{Christopher Jarzynski}
\email{chrisj@lanl.gov}
\affiliation{Theoretical Division, T-13, MSB213, Los Alamos National Laboratory, Los Alamos, New Mexico 87545 }
\begin{abstract}
We consider the adiabatic and quasi-static compression of a dilute classical gas,
confined in a piston and initially equilibrated with a heat bath.
We find that the work performed during this process is described
statistically by a gamma distribution.
We use this result to show that the model satisfies the
non-equilibrium work and fluctuation theorems, but not
the flucutation-dissipation relation.
We discuss the rare but dominant realizations that contribute most
to the exponential average of the work,
and relate our results to potentially universal
work distributions.
\end{abstract}
\preprint{LBNL-59657, LAUR-06-1521}

\maketitle

 \newcommand{\subF}{_{\mathrm{F}}}
  \newcommand{\subR}{_{\mathrm{R}}}

When a system is driven away  an initial state of thermal
equilibrium by a mechanical perturbation, the statistical distribution
of work for that process exhibits universal properties.
In particular, the exponential average of the nonequilibrium work
is related to an equilibrium free energy difference~\cite{Jarzynski1997a, Jarzynski1997b},
\begin{equation}
   \beta \Delta F = -\ln \left \langle e^{-\beta W} \right\rangle = -\ln \int  d W\, \rho(W) e^{-\beta W}
\label{jarzynski}.
   \end{equation}
Furthermore, the work distribution for such a process, 
and the corresponding reversed process, are related by the following  work fluctuation 
theorem~\cite{Crooks1999a, Crooks2000}:
\begin{equation}
\frac{\rho_{\mathrm{F}}(+W)}{\rho_{\mathrm{R}}(-W)} = e^{\beta( W-\Delta F)}
\label{crooks}
\end{equation}
Here, $W$ is the work performed during a given realization of the process~\cite{Crooks1998a};
$\beta$ is the inverse temperature of
a thermal environment with which the system is initially equilibrated;
$\Delta F$ is the free energy difference between two equilibrium states,
both at temperature $\beta^{-1}$,
corresponding to the initial and final values of the external work parameter;
$\rho$ is the work probability distribution;
and the subscripts `$F$' and `$R$' distinguish conjugate forward
and reverse processes, where necessary.
(See Refs.~\cite{Jarzynski1997a,Jarzynski1997b,Crooks1999a,Crooks2000,Crooks1998a,Bustamante05}
for more details and Ref.~\cite{Evans2002} for an overview of related entropy fluctuation relations.)
If we subject the system to a cyclic process
then $\Delta F=0$ and Eq.~\ref{jarzynski} reduces to a result derived by
Bochkov and Kuzovlev\cite{Bochkov1977a, Bochkov1977b}.

The discovery of these relations makes it interesting to find model systems
for which the work distributions can be computed
analytically~\cite{Mazonka1999, Ritort2002, Ritort2004,
LuaGrosberg2005, Lua2005, Bena2005, Dhar2005, Cleuren2005, Bier2005, Imparato2005}.
Exact results have recently been derived for the example of a piston
moving at arbitrary speed against an ideal gas~\cite{LuaGrosberg2005, Lua2005, Bena2005}.
Here we consider the somewhat different case of the quasi-static
compression or expansion of a dilute (but not ideal) classical gas.
This model was suggested in email correspondence to one of us (C.J.)
by Prof.~Seth Putterman, and has also appeared in this setting in a preprint
by Prof.~Jaeyoung Sung~\cite{Sung_condmat0506214v2}.
Using elementary statistical mechanics,
we derive a non-trivial but tractable expression for the work distribution
$\rho(W)$, Eq.~\ref{idealgaswork}, and use this to verify and illustrate
Eqs.~\ref{jarzynski} and \ref{crooks}.

\begin{figure}
\includegraphics{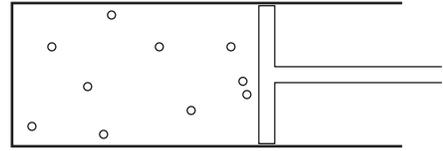} 
\caption{A gas confined to a cylinder with a
controllable piston}
\label{pistonfig}
\end{figure}

\begin{figure*}
\includegraphics{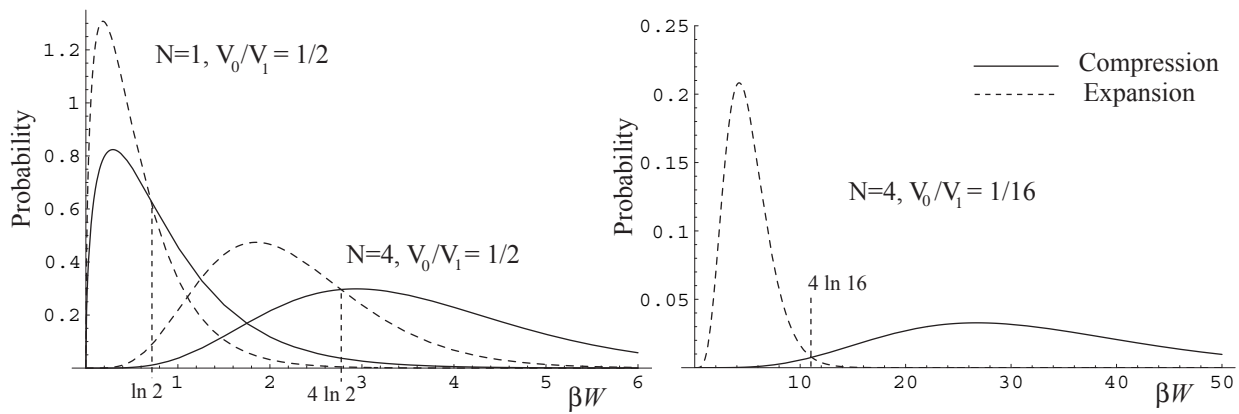} 

\caption{The work probability density, $\rho(W)$, given by 
Eq.~\ref{idealgaswork}.  The solid lines are the work of compression and the dashed lines are the \emph{negative}
work of expansion. Note that the magnitude of the work is greater for compression than expansion.
Each distribution obeys the nonequilibrium work theorem (\ref{jarzynski}) and each compression-expansion pair are 
related by the work fluctuation theorem~(\ref{crooks}).
A direct consequence of the latter, illustrated in the figure, is that the  
corresponding forward and negative reverse work distributions cross at $W=\Delta F$ (\ref{freeEnergyDiff}).
}
\label{workfig}
\end{figure*}

Let us define the model more precisely. Consider the system shown in figure~\ref{pistonfig},
a dilute classical gas confined in a  piston. 
We assume that quantum effects are negligible, that the particles do not have 
any important internal structure, and that they rarely collide with one another.
Specifically, the mean free path between particle-particle collisions is finite (unlike in
Refs.~\cite{LuaGrosberg2005, Lua2005, Bena2005}),
but much greater than the characteristic distance between nearby particles.
Initially, the piston is held fixed and the gas is brought to thermal equilibrium with an external,
infinite heat bath.
The bath is then removed, preventing the further exchange of heat across
the walls of the container.
The piston is then very slowly forced inward, performing work as it compresses
the gas to a new volume.
In the corresponding reverse process we start with the gas at thermal equilibrium
with the final volume of the forward process and then we adiabatically expand
the gas back to the initial volume.

It is useful to define a {\it reference process}, during which the
gas remains in contact with the reservoir, and thus at constant temperature,
as it is compressed reversibly;
$\Delta F$ in Eqs.~\ref{jarzynski}, \ref{crooks}
is the free energy change during this reference process.
By contrast,
during the adiabatic compression described above,
there is a steady rise in the kinetic temperature of the gas.
Thus, although the gas continually self-equilibrates due to
particle-particle collisions,
it is driven away from the isothermal sequence of
equilibrium states defined by the reference process.
 
As a first pass at this model, let us use simple arguments
to verify Eq.~\ref{jarzynski}.
In three spatial dimensions,
the average equilibrium internal energy of a dilute gas
of $N$ identical particles is $\overline{E}= 3N/2\beta$,
and the entropy is given by the Sackur-Tetrode equation,
$S/N = \ln(V/N\Lambda^3) + 5/2$.
Here $V$ is the volume of the box, 
and $\Lambda = \sqrt{\beta h^2/2\pi m}$
is the thermal de Broglie wavelength ($h$ is Planck's constant and $m$ is the particle mass).
The free energy $F=\overline{E}-\beta^{-1}S$ is then
\begin{equation}
\label{F_largeN}
F(\beta,V) = -\frac{N}{\beta}
\left[
\ln\left(\frac{V}{N}\right) +
\frac{3}{2}\ln\left(\frac{2\pi m}{\beta h^2}\right) + 1
\right],
\end{equation}
which satisfies the scaling law (for any $\sigma>0$):
\begin{equation}
\label{scaling}
\sigma F(\sigma\beta,V) = F(\beta,\sigma^{-3/2}V).
\end{equation}

When such a gas is prepared in thermal equilibrium,
as described above,
its energy $E$ can be viewed as a random variable sampled
from the canonical distribution,
\begin{equation}
\label{P}
P(E;\beta,V) = 
\frac{1}{Z(\beta,V)} \, g(E;V) \, e^{-\beta E},
\end{equation}
where $g(E)$ is the density of states and $Z = e^{-\beta F}$ is the partition function.
Since the pressure of a dilute gas is $p = 2E/3V$,
and its energy during an adiabatic process changes by increments $dE = -pdV$,
it follows that the product $VE^{3/2}$ is conserved as we slowly change
the volume from $V_0$ to $V_1$~\cite{Sung_condmat0506214v2}.
The final energy is thus $E_1 = (V_0 /V_1 )^{2/3} E_0$,
and the work performed is
\begin{equation}
\label{we}
W = E_1 -E_0 = \alpha E_0, \quad 
\alpha = \left(\frac{V_0}{V_1}\right)^{2/d} \!\!- 1 ,
\end{equation}
where $d=3$.
Note that $\alpha$ (and therefore $W$), is positive for compression and negative for expansion,
and that $ -1 <\alpha < \infty$.
For expansion the negative work is bounded by the initial kinetic energy.

Defining $q \equiv \alpha + 1= (V_0/V_1)^{2/d}$, we get
\begin{eqnarray*}
- \ln
\Bigl\langle e^{-\beta W}\Bigr\rangle &=&
- \ln
\int dE_0\,P(E_0;\beta,V_0)\,e^{-\beta W(E_0)} \\
=
\label{eq:integral}
- &\ln&
\left[
\frac{1}{Z(\beta,V_0)}
\int dE_0\,g(E_0;V_0)\,e^{-q\beta E_0} \right] \\
=
- &\ln&
\frac{Z(q\beta,V_0)}{Z(\beta,V_0)}
=
q\beta F(q\beta,V_0) - \beta F(\beta,V_0).
\end{eqnarray*}
But $qF(q\beta,V_0) = F(\beta,V_1)$ (Eq.~\ref{scaling}),
hence the right side is simply $\beta\Delta F$, as predicted by Eq.~\ref{jarzynski}.

Although the above analysis is simple, it has its drawbacks.
Eq.~\ref{F_largeN} is a large-$N$ approximation, not an exact result.
Moreover, we have invoked macroscopic, thermodynamic arguments
in deriving Eq.~\ref{we}.
Such arguments are valid when the aim is to describe the
typical behavior of a system,
but become suspect in the present context,
since the average $\langle \exp(-\beta W)\rangle$ is often
dominated by realizations
during which the system behaves very atypically.
Finally, to verify Eq.~\ref{crooks} we must
solve for $\rho(W)$, which requires obtaining the
density of states, $g(E)$.
Therefore, we now proceed with a more careful analysis.
For the sake of generality,
we allow the dimensionality of space to be an arbitrary integer $d>1$,
rather than assuming $d=3$.

The density of states is the derivative of the function $\Phi(E)$,
defined as the number of energy states with energy less than $E$.
For a dilute gas in the classical limit,
\begin{equation}
	\Phi(E;V) = \frac{1}{ h^{2k}} \cdot \frac{V^N}{N!} \cdot \frac{(2 \pi m E)^{k} }{k\,\Gamma(k)}  ,
\label{states}
\end{equation}
where $k=dN/2$,
and $\Gamma(k)$ is the gamma function.
On the right side of Eq.~\ref{states},
the first factor accounts for the quantum graininess of phase space,
the middle factor $V^N/N!$ counts
the number of arrangements of $N$ identical particles in a volume $V$,
and the last factor is the volume of a $dN$-dimensional hypersphere of radius $\sqrt{2 m E}$ in momentum space~\cite{Gibbs1902p93}.
Hence
\begin{equation}
	\label{g}
	g(E;V) = \frac{\partial \Phi}{\partial E}  =
	 \frac{1}{ h^{2k}} 
	 \cdot \frac{V^N}{N!} 
	 \cdot \frac{(2 \pi m )^{k} }{\Gamma(k )}  	 \cdot E^{k-1} .
\end{equation}
The partition function~$Z=\int dE\, g  e^{-\beta E} $ and free energy~$F$ now follow
by direct integration:
\begin{equation}
F(\beta, V) = 
	-\frac{1}{\beta} \ln Z(\beta,V) =
	-\frac{1}{\beta} \ln \left[  	 
	 \frac{V^N}{N!} 
	 \left(\frac{2 \pi m }{ \beta h^2} \right)^{k} 
	 \right] 
\label{freeenergy}
\end{equation}
(We recover Eq.~\ref{F_largeN} with
the approximation $\ln N! \sim N\ln N - N$.)
Eqs.~\ref{P}, \ref{g}, and \ref{freeenergy} together give us
\begin{equation}
	P(E) = \frac{\beta }{\Gamma(k )} (\beta E) ^{k-1}  e^{-\beta E} \; .
	\label{canonical}
\end{equation}


We now solve for $W$ by invoking the quasi-static invariance of
$\Phi(E;V)$~\cite{Hertz1910a,Hertz1910b}.
As discussed in similar contexts
in Refs.~\cite{WatanabeReinhardt1990,Jarzynski1997b,Oberhofer2005,Lua2005},
since the gas continually self-equilibrates,
the value of $\Phi$ remains constant during the process.
From Eq.~\ref{states} we get $\Phi \propto (VE^{d/2})^N$,
therefore $E_1 = (V_0/V_1)^{2/d} E_0$,
which again leads to Eq.~\ref{we}, only now for arbitrary $d>1$.
Thus, using Eq.~\ref{canonical}, we get
\begin{eqnarray}
	\rho(W) &=& \int dE_0 \, P(E_0) \, \delta ( W - \alpha E_0 ) \nonumber \\
	&=&\frac{\beta}{|\alpha| \Gamma(k)} 
		\left( \frac{\beta W}{\alpha}\right)^{k-1}
		e^{-\beta W / \alpha} \,
		\theta (\alpha W),
\label{idealgaswork}
\end{eqnarray}
where the unit step function $\theta$ guarantees that $W$ has the same sign as $\alpha$.
%
%
We see that $\rho(W)$ for adiabatic compression (positive $\alpha$),
and $\rho(-W)$ for adiabatic expansion (negative $\alpha$),
are gamma distributions with shape parameter $k=dN/2$ and scale $s = \vert\alpha\vert/\beta$.
 These distributions are illustrated in figure~\ref{workfig}.
 Note that the work distribution depends on the ratio of the initial and final volumes and not on the absolute volume.

Our result for $\rho(W)$ allows us to verify the fluctuation theorem, Eq.~\ref{crooks}.
Let $\alpha\subF = (V_0/V_1)^{2/d} -1$ and $\alpha\subR = (V_1/V_0)^{2/d} -1$
denote the values of $\alpha$ for the forward ($V_0 \rightarrow V_1$)
and reverse ($V_1 \rightarrow V_0$) processes.
Note that $-\alpha\subR/\alpha\subF = (V_1/V_0)^{2/d}$,
and $\alpha\subF^{-1} +\alpha\subR^{-1} = -1$.
Combining these identities with Eq.~\ref{idealgaswork}, we obtain
 \begin{eqnarray*}
\frac{\rho\subF(+W)}{\rho\subR(-W)} &=& 
\frac{|\alpha\subR|}{|\alpha\subF|}
\left(-\frac{\alpha\subR}{\alpha\subF}\right)^{k -1}
\exp\left[ -\beta W ( \alpha\subF^{-1} +\alpha\subR^{-1}) \right]
\\&=&  \left( \frac{V_1}{V_0} \right)^{N} e^{\beta W}
= e^{\beta( W-\Delta F)},
\end{eqnarray*}
where
\begin{equation}
\label{freeEnergyDiff}
\Delta F = F(\beta,V_1) - F(\beta,V_0) = \frac{N}{\beta}  \ln \frac{V_0}{V_1}
 	 = \frac{1}{\beta} \frac{d N}{2} \ln(1+\alpha)\;,
\end{equation}
by Eqs.~\ref{we}, \ref{freeenergy}.
This confirms Eq.~\ref{crooks}.

The validity of Eq.~\ref{jarzynski} follows immediately from Eq.~\ref{crooks}~\cite{Crooks1999a},
though it can also be verified by the direct evaluation of
$\int \rho(W) e^{-\beta W} \, dW$.
An alternative approach is to 
use a cumulant expansion~\cite{Jarzynski1997a}:
\begin{equation}
\ln \left \langle e^{-\beta W} \right\rangle
=
\sum_{j=1}^\infty (-\beta)^{j} \frac{\omega_j}{j!} ,
\end{equation}
where $\omega_j$ is the $j$'th cumulant of $\rho(W)$.
Using standard properties of the gamma distribution~\cite{Johnson94},
we get
\begin{equation}
\omega_j = \frac{dN}{2}  \left( \frac{\alpha}{\beta} \right)^j (j-1)! 
\end{equation}
hence
\begin{equation}
\label{cumex}
-\ln \left \langle e^{-\beta W} \right\rangle =
-\frac{d N}{2} \,
\sum_{j=1}^\infty \frac{(-\alpha)^{j}}{j} =
\frac{d N}{2} \,
\ln(1+\alpha),
\end{equation}
again confirming Eq.~\ref{jarzynski} for this model (see Eq.~\ref{freeEnergyDiff}).

Truncating this expansion after two terms yields
\begin{equation}
\label{ftr}
\Delta F \approx
\langle W\rangle - \frac{\beta}{2} \sigma_W^2 
= \frac{1}{\beta} \frac{d N}{2} ( \alpha - \alpha^2 ),
\end{equation}
where $\langle W\rangle = \omega_1$ and $\sigma_W^2 = \omega_2$
are the mean and variance of $\rho(W)$.
Eq.~\ref{ftr} is just the fluctuation-dissipation relation of linear response theory.
Naively, we might expect this to be an
excellent approximation for our model,
for either of two reasons.
First, we have assumed a quasi-static process, apparently keeping the system
in the near-equilibrium regime where linear response theory ought to apply.
Second, for $N \gg 1$, the central limit theorem suggests
a Gaussian distribution of work values,
and for a Gaussian only the first two cumulants are non-zero.
However, Eq.~\ref{freeEnergyDiff} reveals that the truncated expansion Eq.~\ref{ftr}
is valid only when $\vert\alpha\vert \ll 1$,
that is, when $V_1 \approx V_0$.
Why does the system not respond linearly for larger $\vert\alpha\vert$?

First, Eq.~\ref{ftr} is valid for small excursions away from the
reversible, {\it isothermal} reference process described earlier.
In our adiabatic process, however,
the kinetic temperature of the gas changes substantially,
hence the system strays far from the reference path,
unless $V_1 \approx V_0$
(From Eq.~\ref{canonical} and $E_1=\alpha E_0$
it follows that the system ends with a canonical distribution
of energies at temperature
$(V_0/V_1)^{2/d} \beta^{-1}$).
Second, while $\rho(W)$ is nearly Gaussian
in the region around its mean,
the average of $\exp({-\beta W})$ is dominated by work values deep in the lower
tail of the distribution, where the central limit theorem does not apply.
Thus we cannot invoke the central limit theorem to throw out the
higher ($j>2$) cumulants;
indeed, the relative sizes of the cumulants are independent of $N$.

Our results also illustrate the Clausius inequality,
\begin{equation}
\langle W\rangle = 
\frac{1}{\beta} \frac{dN}{2} \alpha \ge \frac{1}{\beta} \frac{dN}{2} \ln(1+\alpha) =
\Delta F,
\end{equation}
since $\alpha \ge \ln(1+\alpha)$ for all real $\alpha$.

Assuming many particles,
the initial energy of the gas
is almost always very near to the equilibrium average energy,
$\overline E = dN/2\beta$, implying that the work performed
during a typical realization is
$W^{\rm typ} \approx \alpha \overline E$.
[As a consistency check, Eq.~\ref{idealgaswork} verifies that the peak of
$\rho(W)$ occurs at $\alpha\overline E + {\cal O}(N^{0})$.]
However, the average of $\exp({-\beta W})$ is dominated not by these
typical realizations, but rather by those for which the work falls
near the peak of $\rho(W) \exp({-\beta W})$~\cite{Ritort2004,Jarzynski2006}.
From Eq.~\ref{idealgaswork} we find that this peak occurs at
$W^{\rm dom} \approx \alpha \overline E/q$,
where $q = (V_0 /V_1 )^{2/d}$.
Using Eq.~\ref{we}
we conclude that the dominant realizations
are characterized by energies
\begin{equation}
E_0^{\rm dom} \approx \frac{1}{q} \overline E
\,,\qquad
E_1^{\rm dom} \approx \overline E.
\end{equation}
These are realizations during which the system begins with
a very atypical energy ($\overline E/q$), but ends in a microstate
that is characteristic of thermal equilibrium at temperature $\beta^{-1}$.
This is a generic feature associated with the convergence of
$\langle \exp({-\beta W})\rangle$~\cite{Jarzynski2006}.

As a practical matter, it is often desirable to fit experimental data to an appropriate 
probability distribution. Where the work is smooth and unimodal the gamma
distribution may be a reasonable parametric choice, since it explicitly arises in this physical example,
it obeys the appropriate symmetries, and it can model the skew typical exhibited by work densities.
Indeed in a recent experiment, it was found that the work density could be adequately fit to a Pearson type III distribution,
which is a gamma distribution generalized with a location parameter~\cite{Blickle2005}.
However, the gamma distribution is clearly inadequate for this task in general, since the distribution has a sharp lower bound.
It has long been speculated that the generalized Gumbel (aka generalized Fisher-Tippett) distribution
may provide a universal work distribution, in the sense that the work density for 
many different process may limit towards this general form.
\begin{equation}
P(x) =  \frac{a^a}{\Gamma(a) b}e^{a  b (x-s) + a e^{b(x-s)} } 
\end{equation}
Here, $a$, $b$ and $s$ are real parameters. This distribution 
with positive integer `$a$' arises in extreme value statistics~\cite{Gumbel1958} and has been proposed as a
general model of non-equilibrium fluctuations~\cite{Bramwell2000,Bertin2005}.  
We originally speculated that this may be a plausible general distribution for  work
densities since this is the maximum entropy distribution, from the location-scale 
family, given the mean work and the mean Boltzmann weighted work (Eq.~\ref{jarzynski}).
Moreover, it has recently been shown that the generalized Gumbel provides an excellent fit to work densities 
for various alchemical transformations~\cite{Nanda2005}.
It is therefore interesting to note that for small $b$ the generalized Gumbel approximates the generalized gamma distribution.
Consequentially, the results of this paper are compatible with the universal distribution hypothesis.


\begin{acknowledgments}
This research was 
supported by the Department of Energy, under contracts W-7405-ENG-36 (C.J.) and 
DE-AC02-05CH11231 (G.C.).
C.J. thanks Seth Putterman and Jaeyoung Sung for stimulating correspondence
on this topic.
\end{acknowledgments}

\begin{thebibliography}{32}
\expandafter\ifx\csname natexlab\endcsname\relax\def\natexlab#1{#1}\fi
\expandafter\ifx\csname bibnamefont\endcsname\relax
  \def\bibnamefont#1{#1}\fi
\expandafter\ifx\csname bibfnamefont\endcsname\relax
  \def\bibfnamefont#1{#1}\fi
\expandafter\ifx\csname citenamefont\endcsname\relax
  \def\citenamefont#1{#1}\fi
\expandafter\ifx\csname url\endcsname\relax
  \def\url#1{\texttt{#1}}\fi
\expandafter\ifx\csname urlprefix\endcsname\relax\def\urlprefix{URL }\fi
\providecommand{\bibinfo}[2]{#2}
\providecommand{\eprint}[2][]{\url{#2}}

\bibitem[{\citenamefont{Jarzynski}(1997{\natexlab{a}})}]{Jarzynski1997a}
\bibinfo{author}{\bibfnamefont{C.}~\bibnamefont{Jarzynski}},
  \bibinfo{journal}{Phys. Rev. Lett.} \textbf{\bibinfo{volume}{78}},
  \bibinfo{pages}{2690} (\bibinfo{year}{1997}{\natexlab{a}}).

\bibitem[{\citenamefont{Jarzynski}(1997{\natexlab{b}})}]{Jarzynski1997b}
\bibinfo{author}{\bibfnamefont{C.}~\bibnamefont{Jarzynski}},
  \bibinfo{journal}{Phys. Rev. E} \textbf{\bibinfo{volume}{56}},
  \bibinfo{pages}{5018} (\bibinfo{year}{1997}{\natexlab{b}}).

\bibitem[{\citenamefont{Crooks}(1999)}]{Crooks1999a}
\bibinfo{author}{\bibfnamefont{G.~E.} \bibnamefont{Crooks}},
  \bibinfo{journal}{Phys. Rev. E} \textbf{\bibinfo{volume}{60}},
  \bibinfo{pages}{2721} (\bibinfo{year}{1999}).

\bibitem[{\citenamefont{Crooks}(2000)}]{Crooks2000}
\bibinfo{author}{\bibfnamefont{G.~E.} \bibnamefont{Crooks}},
  \bibinfo{journal}{Phys. Rev. E} \textbf{\bibinfo{volume}{61}},
  \bibinfo{pages}{2361} (\bibinfo{year}{2000}).

\bibitem[{\citenamefont{Crooks}(1998)}]{Crooks1998a}
\bibinfo{author}{\bibfnamefont{G.~E.} \bibnamefont{Crooks}},
  \bibinfo{journal}{J. Stat. Phys.} \textbf{\bibinfo{volume}{90}},
  \bibinfo{pages}{1481} (\bibinfo{year}{1998}).

\bibitem[{\citenamefont{Bustamante et~al.}(2005)\citenamefont{Bustamante,
  Liphardt, and Ritort}}]{Bustamante05}
\bibinfo{author}{\bibfnamefont{C.}~\bibnamefont{Bustamante}},
  \bibinfo{author}{\bibfnamefont{J.}~\bibnamefont{Liphardt}}, \bibnamefont{and}
  \bibinfo{author}{\bibfnamefont{F.}~\bibnamefont{Ritort}},
  \bibinfo{journal}{Physics Today} \textbf{\bibinfo{volume}{58}},
  \bibinfo{pages}{43} (\bibinfo{year}{2005}).

\bibitem[{\citenamefont{Evans and Searles}(2002)}]{Evans2002}
\bibinfo{author}{\bibfnamefont{D.~J.} \bibnamefont{Evans}} \bibnamefont{and}
  \bibinfo{author}{\bibfnamefont{D.~J.} \bibnamefont{Searles}},
  \bibinfo{journal}{Advances in Physics} \textbf{\bibinfo{volume}{51}},
  \bibinfo{pages}{1529} (\bibinfo{year}{2002}).

\bibitem[{\citenamefont{Bochkov and
  Kuzovlev}(1977{\natexlab{a}})}]{Bochkov1977a}
\bibinfo{author}{\bibfnamefont{G.~N.} \bibnamefont{Bochkov}} \bibnamefont{and}
  \bibinfo{author}{\bibfnamefont{Y.~E.} \bibnamefont{Kuzovlev}},
  \bibinfo{journal}{Zh. Eksp. Teor. Fiz.} \textbf{\bibinfo{volume}{72}}
  (\bibinfo{year}{1977}{\natexlab{a}}).

\bibitem[{\citenamefont{Bochkov and
  Kuzovlev}(1977{\natexlab{b}})}]{Bochkov1977b}
\bibinfo{author}{\bibfnamefont{G.~N.} \bibnamefont{Bochkov}} \bibnamefont{and}
  \bibinfo{author}{\bibfnamefont{Y.~E.} \bibnamefont{Kuzovlev}},
  \bibinfo{journal}{Sov. Phys. JETP} \textbf{\bibinfo{volume}{45}}
  (\bibinfo{year}{1977}{\natexlab{b}}).

\bibitem[{\citenamefont{Mazonka and Jarzynski}()}]{Mazonka1999}
\bibinfo{author}{\bibfnamefont{O.}~\bibnamefont{Mazonka}} \bibnamefont{and}
  \bibinfo{author}{\bibfnamefont{C.}~\bibnamefont{Jarzynski}},
  \bibinfo{note}{cond-mat/991212}.

\bibitem[{\citenamefont{Ritort et~al.}(2002)\citenamefont{Ritort, Bustamante,
  and Tinoco}}]{Ritort2002}
\bibinfo{author}{\bibfnamefont{F.}~\bibnamefont{Ritort}},
  \bibinfo{author}{\bibfnamefont{C.}~\bibnamefont{Bustamante}},
  \bibnamefont{and} \bibinfo{author}{\bibfnamefont{I.}~\bibnamefont{Tinoco}},
  \bibinfo{journal}{PNAS} \textbf{\bibinfo{volume}{99}}, \bibinfo{pages}{13544}
  (\bibinfo{year}{2002}).

\bibitem[{\citenamefont{Ritort}(2004)}]{Ritort2004}
\bibinfo{author}{\bibfnamefont{F.}~\bibnamefont{Ritort}}, \bibinfo{journal}{J.
  Stat. Mech.} \textbf{\bibinfo{volume}{2004}}, \bibinfo{pages}{P10016}
  (\bibinfo{year}{2004}).

\bibitem[{\citenamefont{Lua and Grosberg}(2005)}]{LuaGrosberg2005}
\bibinfo{author}{\bibfnamefont{R.~C.} \bibnamefont{Lua}} \bibnamefont{and}
  \bibinfo{author}{\bibfnamefont{A.~Y.} \bibnamefont{Grosberg}},
  \bibinfo{journal}{J. Phys. Chem. B.} \textbf{\bibinfo{volume}{109}},
  \bibinfo{pages}{6805} (\bibinfo{year}{2005}).

\bibitem[{\citenamefont{Lua}(2005)}]{Lua2005}
\bibinfo{author}{\bibfnamefont{R.~C.} \bibnamefont{Lua}}
  (\bibinfo{year}{2005}), \bibinfo{note}{cond-mat/0511302}.

\bibitem[{\citenamefont{Bena et~al.}(2005)\citenamefont{Bena, Van~den Broeck,
  and Kawai}}]{Bena2005}
\bibinfo{author}{\bibfnamefont{I.}~\bibnamefont{Bena}},
  \bibinfo{author}{\bibfnamefont{C.}~\bibnamefont{Van~den Broeck}},
  \bibnamefont{and} \bibinfo{author}{\bibfnamefont{R.}~\bibnamefont{Kawai}},
  \bibinfo{journal}{Euro. Phys. Lett.} \textbf{\bibinfo{volume}{71}},
  \bibinfo{pages}{879} (\bibinfo{year}{2005}).

\bibitem[{\citenamefont{Dhar}(2005)}]{Dhar2005}
\bibinfo{author}{\bibfnamefont{A.}~\bibnamefont{Dhar}}, \bibinfo{journal}{Phys.
  Rev. E} \textbf{\bibinfo{volume}{71}}, \bibinfo{eid}{036126}
  (\bibinfo{year}{2005}).

\bibitem[{\citenamefont{Cleuren et~al.}()\citenamefont{Cleuren, Van~den Broeck,
  and Kawai}}]{Cleuren2005}
\bibinfo{author}{\bibfnamefont{B.}~\bibnamefont{Cleuren}},
  \bibinfo{author}{\bibfnamefont{C.}~\bibnamefont{Van~den Broeck}},
  \bibnamefont{and} \bibinfo{author}{\bibfnamefont{R.}~\bibnamefont{Kawai}},
  \bibinfo{note}{cond-mat/0511653}.

\bibitem[{\citenamefont{Bier}()}]{Bier2005}
\bibinfo{author}{\bibfnamefont{M.}~\bibnamefont{Bier}},
  \bibinfo{note}{cond-mat/0510270}.

\bibitem[{\citenamefont{Imparato and Peliti}(2005)}]{Imparato2005}
\bibinfo{author}{\bibfnamefont{A.}~\bibnamefont{Imparato}} \bibnamefont{and}
  \bibinfo{author}{\bibfnamefont{L.}~\bibnamefont{Peliti}},
  \bibinfo{journal}{Europhys. Lett} \textbf{\bibinfo{volume}{69}},
  \bibinfo{pages}{643} (\bibinfo{year}{2005}).

\bibitem[{\citenamefont{Sung}(2005)}]{Sung_condmat0506214v2}
\bibinfo{author}{\bibfnamefont{J.}~\bibnamefont{Sung}} (\bibinfo{year}{2005}),
  \bibinfo{note}{cond-mat/0506241v2}.

\bibitem[{\citenamefont{Gibbs}(1902)}]{Gibbs1902p93}
\bibinfo{author}{\bibfnamefont{J.~W.} \bibnamefont{Gibbs}},
  \emph{\bibinfo{title}{Elementary Principles in Statistical Mechanics}}
  (\bibinfo{publisher}{Yale University Press}, \bibinfo{year}{1902}),
  \bibinfo{note}{p93}.

\bibitem[{\citenamefont{Hertz}(1910{\natexlab{a}})}]{Hertz1910a}
\bibinfo{author}{\bibfnamefont{P.}~\bibnamefont{Hertz}}, \bibinfo{journal}{Ann.
  Phys. (Leipzig)} \textbf{\bibinfo{volume}{33}}, \bibinfo{pages}{225}
  (\bibinfo{year}{1910}{\natexlab{a}}).

\bibitem[{\citenamefont{Hertz}(1910{\natexlab{b}})}]{Hertz1910b}
\bibinfo{author}{\bibfnamefont{P.}~\bibnamefont{Hertz}}, \bibinfo{journal}{Ann.
  Phys. (Leipzig)} \textbf{\bibinfo{volume}{33}}, \bibinfo{pages}{537}
  (\bibinfo{year}{1910}{\natexlab{b}}).

\bibitem[{\citenamefont{Oberhofer et~al.}(2005)\citenamefont{Oberhofer,
  Dellago, and Geissler}}]{Oberhofer2005}
\bibinfo{author}{\bibfnamefont{H.}~\bibnamefont{Oberhofer}},
  \bibinfo{author}{\bibfnamefont{C.}~\bibnamefont{Dellago}}, \bibnamefont{and}
  \bibinfo{author}{\bibfnamefont{P.~L.} \bibnamefont{Geissler}},
  \bibinfo{journal}{J. Phys. Chem. B} \textbf{\bibinfo{volume}{109}},
  \bibinfo{pages}{6902} (\bibinfo{year}{2005}).

\bibitem[{\citenamefont{Watanabe and Reinhardt}(1990)}]{WatanabeReinhardt1990}
\bibinfo{author}{\bibfnamefont{M.}~\bibnamefont{Watanabe}} \bibnamefont{and}
  \bibinfo{author}{\bibfnamefont{W.~P.} \bibnamefont{Reinhardt}},
  \bibinfo{journal}{Phys. Rev. Lett.} \textbf{\bibinfo{volume}{65}},
  \bibinfo{pages}{3301} (\bibinfo{year}{1990}).

\bibitem[{\citenamefont{Johnson et~al.}(1994)\citenamefont{Johnson, Kotz, and
  Balakrishnan}}]{Johnson94}
\bibinfo{author}{\bibfnamefont{N.~L.} \bibnamefont{Johnson}},
  \bibinfo{author}{\bibfnamefont{S.}~\bibnamefont{Kotz}}, \bibnamefont{and}
  \bibinfo{author}{\bibfnamefont{N.}~\bibnamefont{Balakrishnan}},
  \emph{\bibinfo{title}{Continuous Univariate Distributions}}
  (\bibinfo{publisher}{Wiley and Sons, New York}, \bibinfo{year}{1994}).

\bibitem[{\citenamefont{Jarzynski}(2006)}]{Jarzynski2006}
\bibinfo{author}{\bibfnamefont{C.}~\bibnamefont{Jarzynski}}
  (\bibinfo{year}{2006}), \bibinfo{note}{submitted for publication}.

\bibitem[{\citenamefont{Blickle et~al.}()\citenamefont{Blickle, Speck, Helden,
  Seifert, and Bechinger}}]{Blickle2005}
\bibinfo{author}{\bibfnamefont{V.}~\bibnamefont{Blickle}},
  \bibinfo{author}{\bibfnamefont{T.}~\bibnamefont{Speck}},
  \bibinfo{author}{\bibfnamefont{L.}~\bibnamefont{Helden}},
  \bibinfo{author}{\bibfnamefont{U.}~\bibnamefont{Seifert}}, \bibnamefont{and}
  \bibinfo{author}{\bibfnamefont{C.}~\bibnamefont{Bechinger}},
  \bibinfo{note}{cond-mat/0512461}.

\bibitem[{\citenamefont{Gumbel}(1958)}]{Gumbel1958}
\bibinfo{author}{\bibfnamefont{E.~J.} \bibnamefont{Gumbel}},
  \emph{\bibinfo{title}{Statistics of Extremes}} (\bibinfo{publisher}{Columbia
  University Press, New York}, \bibinfo{year}{1958}).

\bibitem[{\citenamefont{Bramwell et~al.}(2000)\citenamefont{Bramwell,
  Christensen, Fortin, Holdsworth, Jensen, Lise, L\'opez, Nicodemi, Pinton, and
  Sellitto}}]{Bramwell2000}
\bibinfo{author}{\bibfnamefont{S.~T.} \bibnamefont{Bramwell}},
  \bibinfo{author}{\bibfnamefont{K.}~\bibnamefont{Christensen}},
  \bibinfo{author}{\bibfnamefont{J.-Y.} \bibnamefont{Fortin}},
  \bibinfo{author}{\bibfnamefont{P.~C.~W.} \bibnamefont{Holdsworth}},
  \bibinfo{author}{\bibfnamefont{H.~J.} \bibnamefont{Jensen}},
  \bibinfo{author}{\bibfnamefont{S.}~\bibnamefont{Lise}},
  \bibinfo{author}{\bibfnamefont{J.~M.} \bibnamefont{L\'opez}},
  \bibinfo{author}{\bibfnamefont{M.}~\bibnamefont{Nicodemi}},
  \bibinfo{author}{\bibfnamefont{J.-F.} \bibnamefont{Pinton}},
  \bibnamefont{and} \bibinfo{author}{\bibfnamefont{M.}~\bibnamefont{Sellitto}},
  \bibinfo{journal}{Phys. Rev. Lett.} \textbf{\bibinfo{volume}{84}},
  \bibinfo{pages}{3744} (\bibinfo{year}{2000}).

\bibitem[{\citenamefont{Bertin}(2005)}]{Bertin2005}
\bibinfo{author}{\bibfnamefont{E.}~\bibnamefont{Bertin}},
  \bibinfo{journal}{Phys. Rev. Lett.} \textbf{\bibinfo{volume}{95}},
  \bibinfo{pages}{170601} (\bibinfo{year}{2005}).

\bibitem[{\citenamefont{Nanda et~al.}(2005)\citenamefont{Nanda, Lu, and
  Woolf}}]{Nanda2005}
\bibinfo{author}{\bibfnamefont{H.}~\bibnamefont{Nanda}},
  \bibinfo{author}{\bibfnamefont{N.}~\bibnamefont{Lu}}, \bibnamefont{and}
  \bibinfo{author}{\bibfnamefont{T.~B.} \bibnamefont{Woolf}},
  \bibinfo{journal}{J. Chem. Phys.} \textbf{\bibinfo{volume}{122}},
  \bibinfo{pages}{134110} (\bibinfo{year}{2005}).

\end{thebibliography}

\end{document}